%% file: submit.tex
\documentclass[12pt]{article}

\usepackage{lineno}
\usepackage{graphicx}
\usepackage{amsmath}
\usepackage{hhline}
\usepackage{amssymb}
\usepackage{times}
\usepackage{cite}
\usepackage{array}
\usepackage{multirow}
\usepackage{color}

\graphicspath{{./figures/}}

\newif\ifpdf
\ifx\pdfoutput\undefined
    \pdffalse          
\else
    \pdfoutput=1       
    \pdftrue
\fi

\ifpdf
\usepackage{thumbpdf}
\usepackage[pdftex,
        colorlinks=false,
        urlcolor=rltblue,       
        filecolor=rltgreen,     
        linkcolor=rltred,       
        pdftitle={General Search for New Phenomena in $ep$ scattering at HERA},
        pdfauthor={H1},
        pdfsubject={Template file},
        pdfkeywords={High-Energy Physics, Particle Physics},
        pdfpagemode=None,
        bookmarksopen=true]{hyperref}
 
\fi

\definecolor{rltred}{rgb}{0.75,0,0}
\definecolor{rltgreen}{rgb}{0,0.5,0}
\definecolor{rltblue}{rgb}{0,0,0.75}

\newlength{\dinwidth}
\newlength{\dinmargin}
\def\gsim{\,\lower.25ex\hbox{$\scriptstyle\sim$}\kern-1.30ex%
\raise 0.55ex\hbox{$\scriptstyle >$}\,}
\def\lsim{\,\lower.25ex\hbox{$\scriptstyle\sim$}\kern-1.30ex%
\raise 0.55ex\hbox{$\scriptstyle <$}\,}

\newcommand{\Mall}{M_{\mbox{\footnotesize{all}}}}
\newcommand{\pmin}{p_{\mbox{\footnotesize{min}}}}

\newcommand{\ee}{\mbox{$e$-$e$}}
\newcommand{\emu}{\mbox{$e$-$\mu$}}
\newcommand{\ej}{\mbox{$e$-$j$}}
\newcommand{\enp}{\mbox{$e$-$\nu$}}
\newcommand{\epho}{\mbox{$e$-$\gamma$}}
\newcommand{\mumu}{\mbox{$\mu$-$\mu$}}

\newcommand{\muj}{\mbox{$\mu$-$j$}}

\newcommand{\munp}{\mbox{$\mu$-$\nu$}}

\newcommand{\jj}{\mbox{$j$-$j$}}
\newcommand{\jnp}{\mbox{$j$-$\nu$}}
\newcommand{\jpho}{\mbox{$j$-$\gamma$}}
\newcommand{\nppho}{\mbox{$\nu$-$\gamma$}}
\newcommand{\phopho}{\mbox{$\gamma$-$\gamma$}}

\newcommand{\eee}{\mbox{$e$-$e$-$e$}}

\newcommand{\ejj}{\mbox{$e$-$j$-$j$}}
\newcommand{\ejnp}{\mbox{$e$-$j$-$\nu$}}
\newcommand{\ejpho}{\mbox{$e$-$j$-$\gamma$}}

\newcommand{\mujnp}{\mbox{$\mu$-$j$-$\nu$}}

\newcommand{\jjj}{\mbox{$j$-$j$-$j$}}
\newcommand{\jjnp}{\mbox{$j$-$j$-$\nu$}}
\newcommand{\jjpho}{\mbox{$j$-$j$-$\gamma$}}
\newcommand{\jnppho}{\mbox{$j$-$\nu$-$\gamma$}}
\newcommand{\ejjj}{\mbox{$e$-$j$-$j$-$j$}}
\newcommand{\jjjj}{\mbox{$j$-$j$-$j$-$j$}}
\newcommand{\jjjnp}{\mbox{$j$-$j$-$j$-$\nu$}}

\newcommand{\ejjjj}{\mbox{$e$-$j$-$j$-$j$-$j$}}
\newcommand{\jjjjj}{\mbox{$j$-$j$-$j$-$j$-$j$}}

\def\PLB{{\em Phys. Lett.}   {\bf B}}

\begin{document}  

\begin{titlepage}

\noindent

DESY 04--140 \hfill ISSN 0918--9833\\

\vspace*{2cm}

\begin{center}
  \begin{Large}
    {\bf\boldmath  A general search for new phenomena in $ep$ scattering\\ at HERA}\\[2cm]
    H1 Collaboration
  \end{Large}
\end{center}

\vspace{2cm}

\begin{abstract}
  \noindent
  A model-independent search for
  deviations from the Standard Model prediction is performed
  in $e^+ p$ and $e^- p$ collisions at HERA using H1 data
  corresponding to an integrated luminosity of $117$ $\mbox{pb}^{-1}$.
  For the first time all event topologies 
  involving isolated electrons, photons, muons, neutrinos and jets with
  high transverse momenta are investigated in a single analysis.
  Events are assigned to exclusive classes according to their
  final state.
  A statistical algorithm is developed to search for
  deviations from the Standard Model in the distributions of the scalar sum of
  transverse momenta or invariant mass of final state particles and to quantify their significance.
  A good agreement with the Standard Model prediction is observed in most
  of the event classes.
  The most significant deviation is found for a topology containing an isolated muon,
  missing transverse momentum and a jet, consistent with a previously reported observation.

\end{abstract}

\vspace{1.5cm}

\begin{center}
  To be submitted to \PLB 
\end{center}

\end{titlepage}

\begin{flushleft}
  \input{h1auts}
\end{flushleft}
\newpage
\newpage
\pagenumbering{arabic} \setcounter{page}{1}

\section{Introduction}
At HERA electrons\footnote{
  In this paper ``electrons'' refers to both electrons and positrons, if
  not otherwise stated.}
and protons collide at a centre-of-mass energy of up to $319$~GeV. 
These high-energy electron-proton interactions provide a 
testing ground for the Standard Model (SM) complementary to $e^+e^-$ and $p\overline{p}$ scattering. 
It is widely believed that the SM is incomplete and 
that new physics signals may appear below energies of $1$ TeV. 
Many extensions to the SM have been constructed during
the last decades predicting various phenomena which may be visible at high energies 
or large transverse momenta ($P_T$). HERA data have been used to test some 
of these models of new processes by analysing their anticipated experimental signatures
and limits on their parameters have been 
derived~\cite{Kuze:2002vb}.

The approach described in this paper consists of a comprehensive
and generic search for deviations from the SM prediction at large 
transverse momenta. All high $P_T$ final state configurations involving
electrons ($e$), muons ($\mu$), jets ($j$), photons ($\gamma$) or neutrinos ($\nu$) are systematically 
investigated. The analysis covers phase space regions where the SM 
prediction is sufficiently precise to detect anomalies and does not 
rely on assumptions concerning the characteristics of any SM extension. 
Such a model-independent approach might discover 
unexpected manifestations of new physics. Therefore it addresses the important question of whether evidence for new physics  
might still be hidden in the data recorded at collider experiments.
A similar strategy for a model-independent search was previously presented in~\cite{Abbott:2001ke}. 

All final states containing at least two objects ($e$, $\mu$, $j$, $\gamma$, $\nu$) with 
$P_T >$~$20$~GeV in the polar angle\footnote{ 
  The origin of the H1 
  coordinate system is the nominal $ep$ interaction point, with 
  the direction of the proton beam defining the positive 
  $z$-axis (forward region). The transverse momenta are measured 
  in the $xy$ plane. 
  The 
  pseudorapidity $\eta$ is related to the polar 
  angle $\theta$ by $\eta = -\ln \, \tan (\theta/2)$.}
range  $10^\circ < \theta < 140^\circ$ are investigated. 
The complete HERA I data sample ($1994$ -- $2000$) is used, corresponding 
to an integrated luminosity of $117$~pb$^{-1}$.
All selected events are classified into exclusive event classes 
according to the number and types of objects detected in the final state 
(e.g.  \ej, \mujnp, \jjjjj). 
These exclusive event classes ensure a clear
separation of final states and allow an 
unambiguous statistical interpretation of deviations. 
All experimentally accessible combinations of objects have been studied and 
data events are found in $22$ of them.

In a first analysis step the global event yields of the event
classes are compared with the SM expectation.
The distributions of the invariant mass $M_{all}$ and of the scalar sum of transverse momenta
$\sum P_T$ of high $P_T$ final state objects are presented.
New physics may be visible as an excess or a deficit 
in one of these distributions. 
Therefore, in a second step these distributions are systematically investigated using
a dedicated algorithm which locates 
the region with the largest deviation of the data from the SM
prediction. 
The probability of occurrence of such a deviation
is derived, both for each event class individually and globally for all classes combined.
                    
This paper is organised as follows. Section $2$ describes the Standard Model 
processes at HERA and their Monte Carlo simulation. The 
H1 detector, the event selection and measurement procedure
are described in section $3$.
The event yields and distributions for each event class are presented in
section $4$. The search strategy and results are explained in section $5$. 
Section $6$ summarises the paper.

\section{Standard Model processes and Monte Carlo generation}
Several Monte Carlo event 
generators are combined to simulate events for all 
SM processes which have large cross
sections or are expected to be dominant for specific event classes, 
avoiding double-counting. 
All processes 
are generated with an integrated luminosity significantly higher than
that of the data sample and events are passed through a full detector
simulation~\cite{Brun:1987ma}.
At high transverse 
momenta the dominant SM processes are the photoproduction of two jets and 
neutral current (NC) deep-inelastic scattering (DIS).
In the following the abbreviation $X$ represents 
all reaction products other than
the high $P_T$ objects considered.

\paragraph{Photoproduction of jets and photons}
To simulate the direct and resolved photoproduction of jets 
$ep \rightarrow jj X$, prompt photon production $ep \rightarrow \gamma  j X$ 
and the resolved photoproduction of photon pairs 
$ep \rightarrow \gamma \gamma X$, the PYTHIA $6.1$ event 
generator~\cite{Sjostrand:2000wi} is used. Light and heavy flavoured jets 
are generated. The simulation contains the Born level hard scattering
matrix elements and radiative QED corrections. 

\paragraph{Neutral current deep-inelastic scattering}
The Born, QCD Compton and Boson Gluon Fusion matrix elements are used in 
the RAPGAP~\cite{Jung:1993gf} event generator to model NC DIS events. 
The QED radiative effects arising from real photon emission from both the 
incoming and the outgoing electrons are simulated using the 
HERACLES~\cite{Kwiatkowski:1990es} generator. Hence the NC DIS prediction 
contains the processes $ep \rightarrow e j X$, $ ep \rightarrow e j j X$ and 
also models final states with an additional radiated photon. 

\paragraph{Charged current deep-inelastic scattering}
Charged current (CC) DIS events are simulated
using the DJANGO \cite{Schuler:yg} program, which includes first order QED
radiative corrections based on HERACLES. 
This prediction contributes to the final states $ep \rightarrow \nu j X$,
$ep \rightarrow \nu j j X$ and to final states with an additional radiated photon.

\paragraph{QED Compton scattering}
Elastic and quasi-elastic Compton processes $ep \rightarrow e \gamma X$
are simulated with the WABGEN~\cite{Berger:kp} generator. The inelastic 
contribution is already included in the NC DIS RAPGAP sample.

\paragraph{Electroweak production of lepton pairs}
Multi-lepton events ($ee$, $\mu\mu$, $\tau\tau$) are generated with the GRAPE ~\cite{Abe:2000cv}
program, which 
includes all electroweak matrix elements at tree level. 
Multi-lepton production via $\gamma \gamma$, $\gamma Z$, $ZZ$ collisions, 
internal photon conversion and the decay of virtual or real $Z$ bosons 
is considered. Initial and final state QED radiation is included.
The complete hadronic final state is obtained via interfaces to 
PYTHIA and SOPHIA~\cite{Mucke:1999yb} for the inelastic and quasi-elastic regimes, respectively.

\paragraph{W production}
The production of $W$ bosons $ep \rightarrow W X$ and $ep \rightarrow W j X$
is modelled using EPVEC~\cite{Baur:1991pp}. Next-to-leading order QCD 
corrections~\cite{Diener:2002if} are taken into account by reweighting the 
events as a function of the transverse momentum and rapidity of the $W$ 
boson~\cite{Diener:2003df}.

\paragraph{}
Processes with the production of three or more jets, e.g. $ep \rightarrow jjjX$ or 
$ep \rightarrow jjjjX$, are accounted for using leading
logarithmic parton showers as a representation of higher order QCD radiation, with the exception of CC DIS, where the colour-dipole model~\cite{Lonnblad:1992tz} is used.
Hadronisation is modelled using Lund string fragmentation~\cite{Sjostrand:2000wi}.
The prediction of processes with two or more high transverse momentum jets, e.g. $ep \rightarrow jjX$, $ep \rightarrow ejjX$, is scaled by a factor of $1.2$ to normalise the leading order Monte Carlos to next-to-leading 
order QCD calculations~\cite{Adloff:2002au}.

\section{Experimental technique}
\subsection{The H1 detector}
The H1 detector \cite{Abt:1996xv} components relevant to the 
present analysis are briefly described here. Jets, 
photons and electrons are measured with the Liquid
Argon (LAr) calorimeter~\cite{Andrieu:1993kh}, which covers the polar angle range 
$4^\circ < \theta < 154^\circ$ with full azimuthal acceptance. 
Electromagnetic shower energies are measured with a precision of 
$\sigma (E)/E = 12\%/ \sqrt{E/\mbox{GeV}} \oplus 1\%$ and hadronic energies 
with $\sigma (E)/E = 50\%/\sqrt{E/\mbox{GeV}} \oplus 2\%$, as measured in test beams. The central and forward tracking detectors are used to
measure charged particle trajectories, to reconstruct the interaction 
vertex and to supplement the measurement of the hadronic energy. The innermost proportional chamber CIP ($9^\circ <\theta < 171^\circ$) is used to veto charged particles for the identification of photons. The LAr and inner tracking detectors are enclosed in a super-conducting magnetic coil with a strength of $1.15$~T. 
The return yoke of the coil is the outermost part of the detector and is 
equipped with streamer tubes forming the central muon detector 
($4^\circ < \theta < 171^\circ$). It is also used to supplement the 
measurement of hadrons. In the forward region of the detector ($3^\circ < \theta < 17^\circ$) a set of drift chamber layers (the forward muon system) detects muons and, together with an iron toroidal magnet, allows a momentum measurement. The luminosity measurement is based on the Bethe-Heitler process  $ep \rightarrow ep \gamma$, 
where the photon is detected in a calorimeter located
downstream of the interaction point.

The main trigger for events with high transverse momentum is provided 
by the LAr calorimeter. The trigger efficiency is close 
to $100\%$ for events having an electromagnetic deposit in the LAr
(electron or photon) with transverse momentum greater than 
$20$~GeV~\cite{Adloff:2003uh}. Events triggered only by jets have a trigger 
efficiency above $90\%$ for $P_T^{jet}>20$~GeV and nearly $100\%$ for 
$P_T^{jet}>25$~GeV~\cite{matti}.
For events with missing transverse momentum above 20 GeV, determined from an 
imbalance in the transverse momentum measured in the calorimeter, the trigger efficiency is 
$\sim$ $90\%$.
The muon trigger is based on single muon signatures from the central muon detector, which are combined with signals from the central tracking detector.
The trigger efficiency for di-muon events is
about $70\%$~\cite{Aktas:2003sz}.

\subsection{Event selection}
At HERA electrons or positrons with an energy 
of $27.6$ GeV collide with protons at an energy of $920$ GeV resulting in a
centre-of-mass energy of $\sqrt{s}$ = $319$~GeV. Before $1998$ 
the proton energy was $820$ GeV resulting in a centre-of-mass 
energy of $\sqrt{s}$ = $301$~GeV. The event sample studied consists of the 
full $1994$--$2000$ HERA I data set. It corresponds to an integrated luminosity 
of $36.4$ $\mbox{pb}^{-1}$ in $e^+p$ scattering  at $\sqrt{s}=301$~GeV 
and $13.8$ $\mbox{pb}^{-1}$ in $e^-p$ scattering and $66.4$ $\mbox{pb}^{-1}$ in $e^+p$ scattering at 
$\sqrt{s}=319$~GeV.  

The data selection requires at least one isolated electromagnetic cluster, 
jet or muon  to be found in the detector acceptance. 
Energy deposits in the calorimeters and tracks in the inner tracking system
are used to
form combined cluster-track objects, from which the hadronic energy is
reconstructed.
To reduce background it is demanded that the event 
vertex be reconstructed within $35$~cm 
in $z$ of the nominal position\footnote{
  This is not required for the event classes containing 
  only photons or photons and a neutrino.} 
and that $\sum_i \left(E_i-P_{z,i}\right)< 75$~GeV, where $E_i$ is the 
particle energy and 
$P_{z,i}$ is the $z$ component of the particle momentum.
Here, the index $i$ runs over all hadronic energy deposits, electromagnetic clusters
and muons found in the event. Due to energy-momentum conservation events are expected to have a value of \mbox{$\sum_i \left(E_i-P_{z,i}\right)=55.2$}~GeV, twice the electron beam energy,
if only longitudinal
momentum along the proton beam direction is unmeasured. 
Events with topologies 
typical of cosmic ray and beam-induced background are rejected~\cite{Zhang:2000hb}.
Moreover, the timing of the event is required to coincide with that of the $ep$ bunch crossing.

The identification criteria for each type of particle are based on 
those applied in previous analyses of specific 
final states~\cite{Adloff:2002au,Adloff:2003uh,Aktas:2003sz,Andreev:2003pm}. 
Additional requirements are chosen to ensure an unambiguous 
identification of particles, whilst retaining high efficiencies. 
The following paragraphs describe the identification criteria for the different objects and give the identification efficiencies for the kinematic region considered in the analysis.
 
\paragraph{Electron identification}
The electron identification is based on the measurement 
of a compact and isolated electromagnetic shower in the LAr calorimeter. 
The hadronic energy within a distance in the pseudorapidity-azimuth
($\eta-\phi$) plane $R=\sqrt{(\Delta \eta)^2+(\Delta \phi)^2}<0.75$ around the 
electron is required to be below $2.5\%$ of the electron energy.
This calorimetric electron identification is complemented by 
tracking conditions. A high quality 
track is required to geometrically match the electromagnetic cluster within a 
distance of closest approach to the cluster centre-of-gravity of $12$~cm. 
No other good track is allowed within $R<0.5$ around the electron direction. 
In the central region ($20^\circ < \theta <  140^\circ$) the distance between the first 
measured point in the central drift chambers and the beam axis is required to 
be below $30$~cm in order to reject photons that convert late in the central 
tracker material. In addition, the transverse 
momentum measured from the associated track $P_T^{e_{tk}}$ is required 
to match the calorimetric measurement $P_T^e$ with 
$1/P_T^{e_{tk}} - 1/P_T^e < 0.02$~GeV$^{-1}$. 
In the region not fully covered by the central drift chambers 
($10^\circ < \theta <  37^\circ$) a wider isolation 
cone of $R=1$ is required to reduce the contribution of fake electrons from hadrons. 
In this forward region the identification is completed by the requirement of associated hits 
in the CIP, 
which reduces the contamination from neutral particles showering in the material of the forward region.
The resulting electron finding efficiency 
is $85$\% in the central region and $70\%$ in the forward region.

\paragraph{Photon identification}
The photon identification relies on the measurement of an 
electromagnetic shower and on the same calorimetric isolation criteria against 
hadrons as for the electron identification.
In addition, photons are required to be separated from jets with $P_T^{jet} > 5$~GeV by a distance of $R>1$ to the jet axis. 
Vetoes are applied on any charged track pointing to the electromagnetic 
cluster. No track should be present
with a distance of closest approach to the cluster below $24$~cm or 
within $R<0.5$. An additional veto on any hits in the CIP
is applied.
The resulting photon identification efficiency as derived using 
elastic QED Compton events is $90\%$.

\paragraph{Muon identification}
The muon identification is based on a track in the forward muon
system or in the inner tracking systems associated with
a track segment or an energy deposit in the central muon detector~\cite{Andreev:2003pm}. The muon momentum is measured from the track 
curvature in the toroidal or solenoidal magnetic fields. A muon candidate 
should have no more than $8$~GeV deposited in the LAr calorimeter in a 
cylinder of radius $0.5$ in ($\eta-\phi$) space, centred on the muon track direction. In di-muon events, the requirement of an opening angle between 
the two muons smaller than $165^\circ$ discards  
cosmic ray background. Beam halo muons are rejected by requiring 
that the muons originate from the event vertex. 
Finally, misidentified  hadrons 
are almost completely suppressed by requiring that the  muon candidate is 
separated from the closest jet with 
$P_T^{jet}>5$~GeV by $R>1$. The efficiency to 
identify muons is greater than $90\%$~\cite{Andreev:2003pm}.

\paragraph{Jet identification}
Jets are defined using the 
inclusive $k_{\bot}$ algorithm \cite{Ellis:1993tq,Catani:1993hr}. 
The algorithm is applied in the laboratory frame with a separation parameter
of $1$ and using a $P_T$ weighted recombination scheme \cite{Ellis:1993tq} in which the jets are
treated as massless. 
The jet algorithm is run on all combined cluster-track objects not previously 
identified as electron or  photon candidates. The scattered electron may fake a jet.
This effect is important for multi-jet events, especially 
at high transverse momenta. To reject these fake jets, the first radial moment of the jet transverse energy \cite{Giele:1997hd,frising} is required to be greater than $0.02$ and the quantity
$M^{jet}/P_T^{jet}$ must be greater than $0.1$~\cite{Adloff:2002au,frising}. 
The invariant mass $M^{jet}$ is obtained using the four-vector of all objects belonging to the jet. If the fraction of the jet energy contained in the 
electromagnetic part of the LAr calorimeter is greater than $0.9$, the above
criteria are tightened to $0.04$ and $0.15$, respectively. The jet selection efficiency is $97\%$.

\paragraph{Neutrino identification}
A neutrino candidate is defined in events with missing transverse momentum
above $20$~GeV. The missing momentum is derived from all identified particles and
energy deposits in the event. Fake missing transverse momentum may also arise from the
mismeasurement of an identified object. This effect is reduced
by requiring that the neutrino\footnote{
  The four-vector of the neutrino is calculated under 
  the assumption $\sum_i \left(E_i-P_{z,i}\right) + \left(E_\nu -P_{z,\nu}\right) =55.2$~GeV.} be isolated from 
all identified objects with a transverse momentum above $20$~GeV.
Requiring $\sum_i \left(E_i-P_{z,i}\right)<48$~GeV discards neutrino 
candidates from NC processes where the missing transverse momentum is caused by energy leakage in the 
forward region. 
If exactly one electron or muon object is found, a neutrino object is only assigned to an event 
if $\Delta \phi(l-X_{tot})<170^\circ$, where $\Delta \phi(l-X_{tot})$ is the 
separation in azimuthal angle between the lepton $l$ and the direction of the system $X_{tot}$ built of all hadronic energies. 
 
\paragraph{Event classification}
The common phase space for electrons, photons, muons and jets is defined 
by $10^\circ<\theta<140^\circ$ and $P_T > 20$~GeV. The  neutrino phase space 
is defined by missing transverse momentum above $20$~GeV and  
$\sum_i \left(E_i-P_{z,i}\right)<48$~GeV. These values are chosen to retain a high 
selection and trigger efficiency. All particles with $P_T > 20$~GeV, 
including the neutrino defined by its reconstructed four-vector, are required 
to be isolated compared with each other by a minimum distance $R$ of one unit in the 
$\eta-\phi$ plane. The events are 
classified, depending on the number and types of objects, into
exclusive event classes. Events with an isolated calorimetric 
object in the considered phase space which is not identified as a photon, 
electron or jet are discarded from the analysis in order to minimise wrong 
classifications.  

Based on these identification criteria, purities can be 
derived for each event class with a sizeable SM expectation. Purity is defined as the ratio of SM events reconstructed 
in the event class in which they are generated to the total number of 
reconstructed events in this class. 
Most purities are found to be above $60\%$ and they 
are close to $100\%$ for the 
\jj, \ej, \jnp~and \mumu~event classes.

\subsection{Systematic uncertainties}
This section describes the sources of experimental and theoretical systematic 
uncertainties considered. Experimental systematic uncertainties arising from 
the measurement of the objects are listed in table~\ref{tab:objectunc} (for more details see ~\cite{matti,frising,martin}).
  
\begin{itemize}
\item  
  The electromagnetic energy scale uncertainty varies between $0.7\%$ and $3\%$ depending on the particle's impact point on the LAr calorimeter surface~\cite{Adloff:2003uh}. 
  The polar angular measurement uncertainty of electromagnetic clusters varies depending 
  on $\theta$ between $1$ and $3$~mrad~\cite{Adloff:2003uh}.
  The identification of electron and photon candidates 
  depends on the tracking efficiency, which is known with a precision ranging from $2\%$ for polar 
  angles above $37^\circ$ to $7\%$ in the forward region. 

\item
  The hadronic energy scale of the LAr calorimeter is known to $2\%$. 
  The uncertainty on the jet polar angle determination is $5$ mrad 
  for $\theta<30^\circ$ and $10$ mrad for $\theta>30^\circ$.
\item
  The uncertainty on the transverse momentum measurement for muons amounts to $5\%$. 
  The uncertainty on the polar angle is $3$~mrad. The muon identification efficiency is known with a precision of 5\%.
\item
  The trigger uncertainties for each class are determined by 
  the object with the highest trigger efficiency. The uncertainty on the
  trigger efficiency is estimated to be $3\%$ if the event is triggered
  by a jet or missing transverse momentum and $5\%$ if it is triggered by a muon. For electrons and photons the 
  uncertainty on the trigger efficiency is negligible.
\item
  The uncertainty in the integrated luminosity results in an
  overall normalisation error of $1.5\%$.
\item
     The uncertainty in the reconstruction of $ \sum_i \left(E_i-P_{z,i}\right)$ and the missing $P_T$
  for the neutrino classification are obtained by propagation of the
  systematic errors for other objects.
\end{itemize}

Depending on the generator level
production process, different theoretical uncertainties are used as listed in table~\ref{tab:modelunc}. The errors attributed to 
the predictions for $ep\rightarrow jj X$, $ep \rightarrow j\gamma X$,
$ep\rightarrow j\nu X$, $ep \rightarrow jeX$, 
$ep\rightarrow jj\nu X$, $ep \rightarrow jjeX$ 
and $W$ production include uncertainties in the parton distribution functions and those
due to missing higher order corrections~\cite{Adloff:2002au,Andreev:2003pm,frising,martin}. 
The error attributed to $ep \rightarrow \mu\mu X$ and $ep \rightarrow ee X$ results 
mainly from the limited knowledge of the proton 
structure~\cite{Aktas:2003sz,Aktas:2003jg}. 
The error on the 
QED Compton cross section is estimated to be $5$\% for elastic
and 10\% for inelastic production. An additional theoretical error of 
$20\%$ is applied for each jet produced by parton shower processes 
(e.g. 20\% for the \jjj~event class).
An uncertainty of $50\%$ is added to the prediction for NC DIS 
events with missing transverse 
momentum above $20$~GeV and a high $P_T$ electron. This uncertainty is 
estimated by a comparison of the missing transverse momentum distribution 
between NC DIS events with a low $P_T$ electron ($P_T<20$~GeV) and the SM prediction.

All systematic errors are added in quadrature and are assigned to the SM predictions. For example, the resulting total uncertainties for \ej~events are $10\%$ and $35\%$ at low and high invariant mass $M_{all}$, respectively. In the \jj~event 
class the errors are typically $20\%$ and reach $40\%-50\%$ for $M_{all}$ and
$\sum P_T$ values around $250$~GeV.

\section{Event yields}
\begin{figure}[p]
  \center
  \includegraphics[width=0.565\textheight]{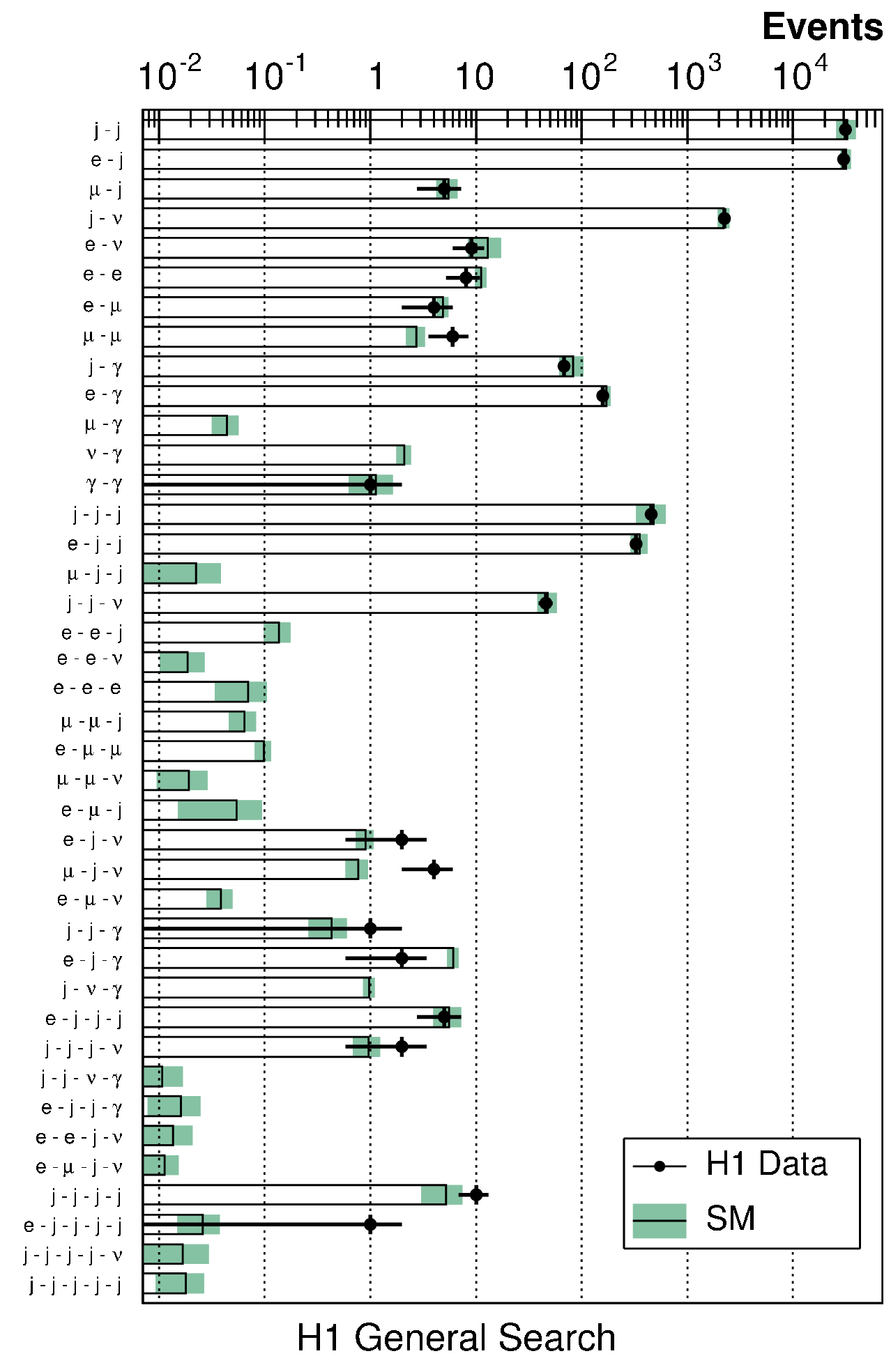}
  \caption{The data and the SM expectation for all event classes 
    with a SM expectation greater than $0.01$ events.
    The analysed data sample corresponds to an integrated luminosity of 117~pb$^{-1}$. 
    The error bands on the predictions include model uncertainties and 
    experimental systematic errors added in quadrature.
  }
  \label{fig:summaryplot}
\end{figure}

\begin{figure}[p]
  \includegraphics[width=\textwidth]{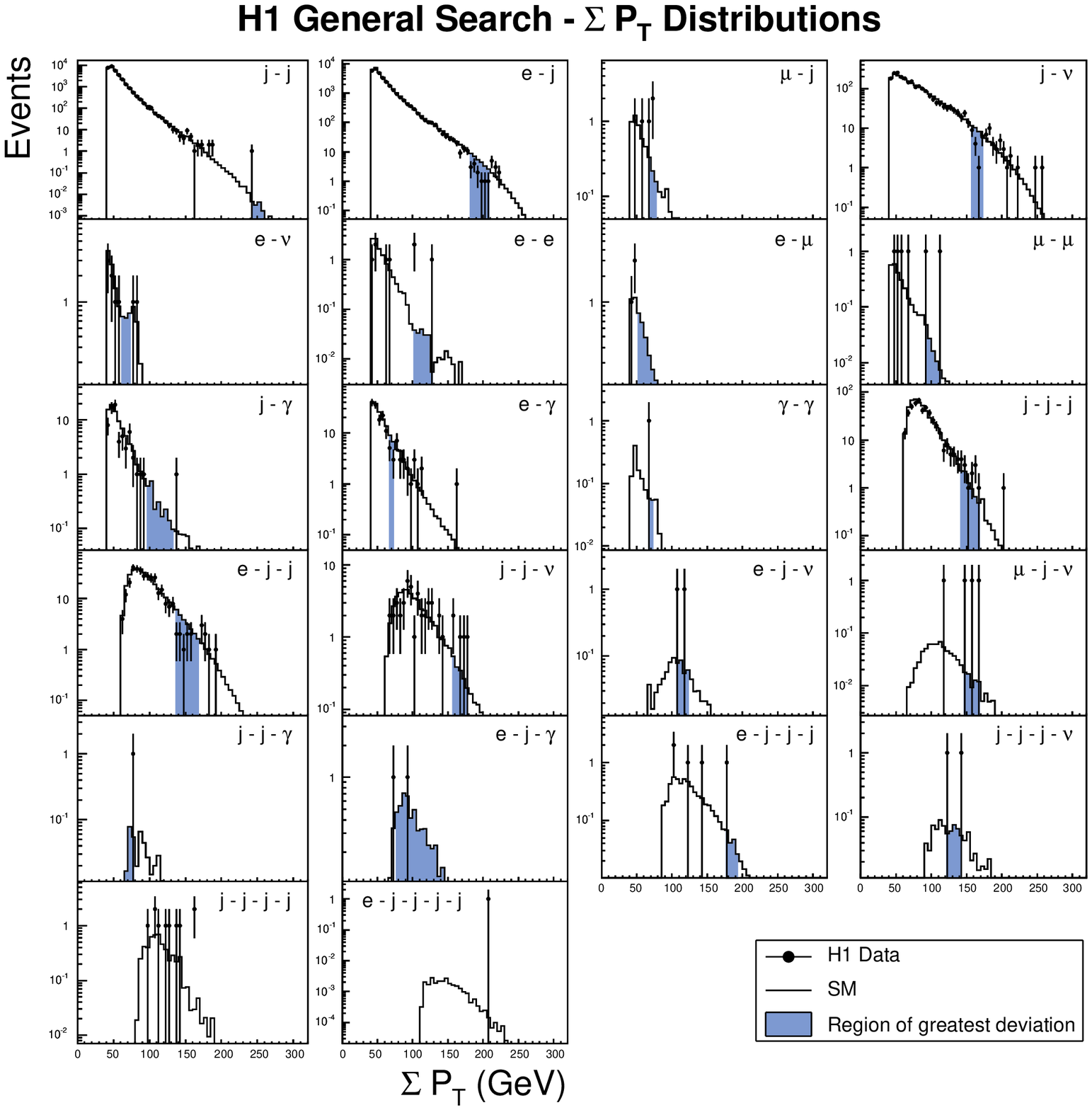}
  \caption{The number of data events and the SM expectation as a 
    function of $\sum P_T$ for classes with at least one event. 
    The shaded areas show the regions of greatest deviation 
    chosen by the search algorithm. No search is performed for the \jjjj~and \ejjjj~ classes.}
  \label{fig:1}
\end{figure}

\begin{figure}[p]
  \includegraphics[width=\textwidth]{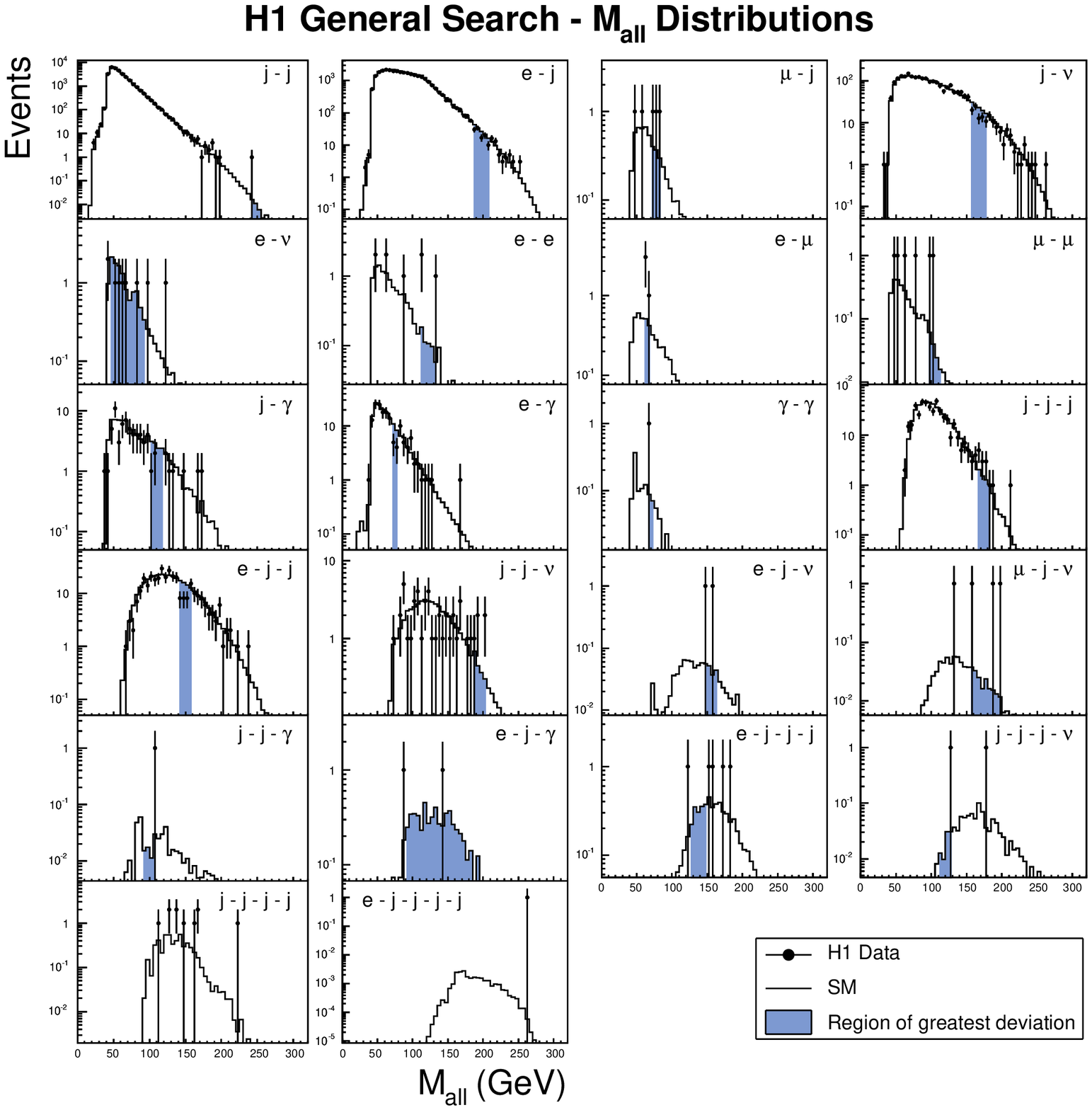}
  \caption{The number of data events and the SM expectation as a
    function of $\Mall$ for event classes with at least one event. 
    The shaded areas show the regions of greatest deviation 
    chosen by the search algorithm. No search is performed for the \jjjj~and \ejjjj~ classes.}
  \label{fig:2}
\end{figure}

All possible event classes with at least two objects are investigated\footnote{The \munp~ event class is discarded from the present analysis.
It is dominated by low transverse energy photoproduction 
events in which a poorly reconstructed muon
gives rise to missing transverse momentum, which fakes the
neutrino signature.}. 
The event yields subdivided 
into event classes are presented for the data and SM expectation
in figure~\ref{fig:summaryplot}. All event classes with
a SM expectation greater than $0.01$ events are shown. No other event class
 contains data events. The distributions of the scalar sum of transverse 
momenta $\sum P_T$ and of the invariant mass $M_{all}$ of all objects are
presented in figures \ref{fig:1} and \ref{fig:2} for classes with at least one event. 

The dominant high $P_T$ processes at HERA, i.e. photoproduction of 
jets, NC and CC DIS, dominate in the \jj, \ej~and \jnp~event 
classes, respectively. Events are observed with $\sum P_T$ and $M_{all}$ values as large as $250$~GeV. 
A good description of the
data spectra by the prediction is observed.
The prediction for the event classes \jjpho~ and \ejpho~is dominated by photoproduction and NC DIS processes with the radiation of a 
photon, respectively. 
There is good agreement between the data and the prediction.
No event is observed in the radiative CC classes \nppho~and \jnppho, consistent
with the expectation of $2.1$ $\pm$ $0.3$ and $1.0 \pm 0.1$, respectively.
The \jjj, \ejj, \ejjj, \jjnp~and \jjjnp~event classes correspond to 
processes with additional jet production due to higher order QCD radiation. 
The yields of these event classes are also well described by the SM prediction. 

The \epho~event class is dominated by QED Compton scattering processes and $\sum P_T$ and $M_{all}$ values up to $160$~GeV are 
measured. A good agreement with the SM is observed.
The prompt photon \jpho~event class extends up to $M_{all}$ $\sim$ $150$ GeV and is well described by the prediction. The purity in this class is moderate ($40-50\%$) due to the high background from misidentified electrons in NC DIS. Backgrounds where hadrons are misidentified as photons are small. 
One event is observed in the \phopho~event class for an expectation of 1.1 $\pm$ 0.5, which is dominated by the $ep \rightarrow e \gamma X$ process, where the electron 
is misidentified.
Contributions of 
higher order QED processes, which could lead to two high transverse momentum  
photons, are negligible.

Lepton pair production from $\gamma \gamma$ processes dominates in event classes with several leptons. 
The \ee~event class contains $8$ events for an 
expectation of $11.2 \pm 1.4$. In this channel, a discrepancy with the SM expectation was previously reported for high masses by the H1 
collaboration~\cite{Aktas:2003jg}. All multi-electron events mentioned in~\cite{Aktas:2003jg}
and located in the phase space of this analysis are found. 
In the region $M_{all}>100$~GeV, $3$ events are observed and 
$1.16\pm 0.25  $ are expected. The higher SM prediction compared 
with the prediction of $0.3$ in~\cite{Aktas:2003jg} 
is due to the extended polar angle range in the forward region. This leads to 
an additional $\approx 0.4$ background events
from fake electrons and $\approx 0.4$ events from 
the $ep\rightarrow eeX$ processes. 
The \eee~class contains no events. None of the  
tri-electron events of~\cite{Aktas:2003jg} are selected here due to the requirement of high transverse momentum.
The predictions for the \emu~and \mumu~event classes are dominated by muon pair 
production from two-photon reactions. The \emu~event class is 
populated if the scattered electron and only one of the muons are selected.
In the \emu~class, $4$ events are observed 
compared with an expectation of $4.8 \pm 0.6$. A slight excess is observed in the \mumu~event class where $6$ events are found and $2.7 \pm 0.6$ are expected.
Muon pair production processes also contribute $\approx$  $85\%$ in
the \muj~event class, where a good agreement is found. 
In the \emu, \mumu~and \muj~event classes the  $\sum P_T$ and $M_{all}$ values 
of the data lie between $50$ and $100$~GeV.

The prediction for the event classes \mujnp~and~\ejnp~consists mainly of high $P_T$ $W$ production with a subsequent leptonic 
decay. 
A discrepancy between the data and the SM expectation is observed 
in the \mujnp~event class, where $4$ events are observed for 
an expectation of $0.8\pm 0.2$. The $\sum P_T$ values reach $170$~GeV 
and the $M_{all}$ values $200$~GeV. 
In this event class 
less than $0.002$ background events are expected from the photoproduction of jets via QCD processes.
Such a deviation was previously reported in~\cite{Andreev:2003pm} 
and will be further discussed in Section 5. 
In the \ejnp~event class $2$ data events are observed for 
an expectation of $0.9 \pm 0.2$. Some of the \ejnp~events mentioned in~\cite{Andreev:2003pm} have an electron with a transverse momentum 
below $20$~GeV and are therefore not selected as \ejnp~events 
in the present analysis.
The event topology
\enp~is also expected to contain events arising from $W$ production together with background from NC DIS.
In the \enp~event class, $9$ data events are observed compared with 
an expectation of $12.9 \pm 4.5$.

\begin{figure}
  \begin{center} 
    \includegraphics[width=0.75\textwidth]{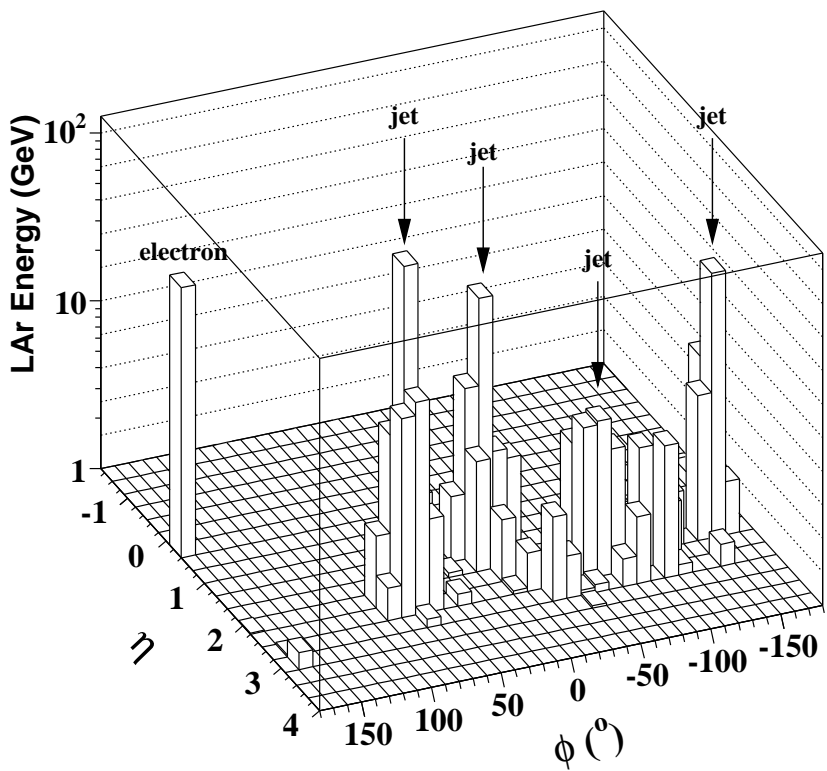}\\
    \caption{The calorimetric energy deposits of the $\ejjjj$ event as a function of $\eta$ and $\phi$.}
    \label{fig:evdis}
  \end{center}
\end{figure}

A slight excess of the data compared 
with the prediction is observed in the \jjjj~event class, with $10$ data events observed and $5.2$ $\pm$ $2.2$ expected. 
One event is observed in the \ejjjj~event class, to be compared with an 
expectation of $0.026\pm 0.011$. 
This event has a $\sum P_T$ of $207$~GeV and 
an invariant mass $M_{all}$ of $262$~GeV.
The NC DIS expectation for $M_{all}>260$~GeV is $(9 \pm 6) \cdot
10^{-5}$ as derived using RAPGAP.
The energy flow of the event in the $\eta - \phi$ view is presented in figure~\ref{fig:evdis}.
The NC DIS and photoproduction SM predictions have been tested 
using a sample of \jjjj~ events with $P_T^{jet} >$~$15$~GeV and \ejjjj~events with $P_T^e > 10$~GeV and $P_T^{jet} > 5$~GeV. 
An adequate 
description of the $\sum  P_T$ 
and $M_{all}$ distributions of the data is obtained within the quoted SM uncertainties.
Since the NC DIS prediction for $M_{all} > 260$ GeV is only of order 0.001 fb, rare SM processes not considered in this analysis such as $W$ pair
 production may be dominant in this kinematic domain. 

No events are found in any other event class, 
in agreement with the SM expectation (see figure~\ref{fig:summaryplot}).

\section{Search for deviations from the Standard Model}
\subsection{Search algorithm and strategy}
In order to quantify the level of agreement between 
the data and the SM expectation and to identify regions of possible 
deviations, a new search algorithm is developed. 
Detailed studies have shown that $M_{all}$ and $\sum P_T$ have a large sensitivity to new physics (see appendix and \cite{martin}). The algorithm
described in the following locates the region of largest deviation of the data from the SM in these 
distributions. The calculation 
of the significance of this deviation is inspired 
by~\cite{Abbott:2001ke}. 

\paragraph{Definition of regions}
A region in the $\sum P_T$ and $M_{all}$ distributions is defined as a set of connected histogram bins\footnote{
  In order to minimise binning effects, a bin size
smaller than or comparable with the resolution of the studied quantity is 
used.
A 5 GeV bin size is used for all distributions. 
 Further reduction of the bin size has a negligible
  effect on the results.} with a size of at least twice the resolution.
All possible regions of any width and at any position in the histograms are considered. 
The number of data events ($N_{obs}$), the SM 
expectation ($N_{SM}$) and its total systematic uncertainty ($\delta N_{SM}$) are calculated for each region.

\paragraph{Determination of the most interesting region}
A statistical estimator $p$ is defined to judge which region is of 
most interest. This estimator is derived from the convolution of the
Poisson probability density function (pdf) to account for statistical 
errors with a Gaussian pdf, $G(b;N_{SM},\delta N_{SM})$, with mean $N_{SM}$ and width $\delta N_{SM}$, to include the effect of 
non negligible systematic uncertainties. The estimator
is defined via 
\begin{equation*}
  p  = \left\{ \begin{array}{ll}
      A \int\limits_0^{\infty} db  \, G(b;N_{SM},\delta N_{SM}) \, \sum\limits_{i=N_{obs}}^{\infty} \frac{e^{-b}
        b^i} {i!} & \textrm{ if } N_{obs} \ge N_{SM} \\
      A \int\limits_0^{\infty} db  \, G(b;N_{SM},\delta N_{SM}) \,\,\,\, \sum\limits_{i=0}^{N_{obs}} \frac{e^{-b}
        b^i} {i!} & \textrm{ if } N_{obs} < N_{SM} 
    \end{array} \right.
\end{equation*}
\begin{equation*}
    \textrm{ with } \,\,\,\, A= 1 / \left[ \int\limits_0^{\infty} db  \, G(b;N_{SM},\delta
  N_{SM}) \, \sum\limits_{i=0}^{\infty} \frac{e^{-b} b^i} {i!}  \right].
\end{equation*}
The factor $A$ ensures normalisation to unity. If $G$ is replaced 
by a Dirac delta function $\delta(b-N_{SM})$ the estimator $p$ becomes the usual 
Poisson probability. The value of $p$ gives an estimate of the probability of a fluctuation of the SM expectation upwards (downwards) to at least (at most) the observed number of data events in the region considered.
The region of greatest deviation 
is the region having the smallest $p$-value, $\pmin$. 
Such a method is able to find narrow resonances and single outstanding events 
as well as signals spread over large regions of phase space in
distributions of any shape~\cite{martin}.

\paragraph{Significance per event class}
The probability that a fluctuation with a $p$-value at least as
small as $\pmin$ occurs
anywhere in the distribution is estimated
using the following method. 
Many independent hypothetical data histograms are made by filling each bin
with an event number diced
according to the pdfs of the SM expectation
(again a convolution of Poisson and Gaussian pdfs). 
For each of 
these hypothetical data histograms the algorithm is run to find the region of 
greatest deviation and the corresponding $\pmin^{SM}$ is calculated. 
The probability $\hat{P}$ is then defined as the 
fraction of hypothetical data histograms with a $\pmin^{SM}$ equal to or smaller 
than the $\pmin$ value obtained from the data.
$\hat{P}$ is a measure of the statistical significance of the deviation observed in the data.
If the event classes are exclusive, the $\hat{P}$ values can be used 
to compare the results of different event classes. 
Depending on the final state, a $\pmin$-value of $5.7 \cdot 10^{-7}$ (``$5\sigma$'') corresponds 
to a value of $- \log_{10}{\hat{P}}$ between $5$ and $6$.


\paragraph{Global significance}
The overall degree of agreement with the SM can be further quantified by 
taking into account the large number of event 
classes studied in this analysis.
The probability of observing an event class with a given $\hat{P}$ value
or smaller can be calculated with 
Monte Carlo (MC) experiments. 
A MC experiment is defined as
a set of hypothetical data histograms
(either in $M_{all}$ or in
$\sum P_T$) following 
the SM expectation with an integrated luminosity of 117 pb$^{-1}$,
on which the complete search algorithm and statistical analysis are
applied as for data. This procedure is repeated many times.
The expectation for the $\hat{P}$ values 
observed in the data is then given by the 
distribution of $\hat{P}^{SM}$ values obtained from all MC experiments.
The probability to find a $\hat{P}$ value smaller than the
minimum observed
in the data can 
thus be calculated and quantifies
the global significance of the observed deviation.

\subsection{Search results}
The final $\hat{P}$ values obtained for event classes having 
at least one observed event are 
summarised in table~\ref{tab:phatmass}.
The regions selected by the algorithm
are presented in figures \ref{fig:1} and \ref{fig:2}. 

The most significant 
deviation of the analysis is found in the \mujnp~event class. This class has  $\hat{P}$ values of $0.010$ ($M_{all}$)
and $0.001$ ($\sum P_T$). The mass region ($155<M_{all}<200$~GeV) contains $3$ data events for an 
expectation of $0.19\pm 0.05$. In the chosen $\sum P_T$ region ($145<\sum P_T<170$~GeV) three data events are 
found while only $0.07\pm 0.03$ are expected. This event topology was 
studied in ~\cite{Andreev:2003pm} where this deviation at high $P_T$ was
already reported.


A $\hat{P}$ value of $0.019$ is found in the \ee~event class in a region at high transverse momenta, $100<\sum P_T<130$~GeV
where 3 events are observed 
for an expectation of $0.18\pm0.08$. The deviation is less prominent in the region selected in the invariant mass distribution
due to a higher background from NC DIS events. 
This corresponds to the
excess of data events also identified in~\cite{Aktas:2003jg}.

A deficit is observed in the \ej~event class in the $\sum P_T$ distribution in the
region
$180 < \sum P_T$~$<$~$210$~GeV.
For a SM expectation of $31.2\pm 5.0$ only $12$ data events are observed. 
The derived $\hat{P}$ value is  $0.021$.

Due to the uncertainties in the SM prediction in the \jjjj~and \ejjjj~event 
classes at the highest $M_{all}$ and $\sum P_T$, where data events are 
observed (see section 4), no reliable $\hat{P}$ values can be 
calculated for these classes. 
Consequently, these event classes are not taken into account to determine the overall degree of agreement between the data and the SM.

The $\hat{P}$ values for event classes with no data event observed and a small
SM expectation are $1$. This remains the case if an additional
contribution is added from a possible further rare process not included
here. Such classes can thus be considered in the calculation
of the global significance.

\begin{figure}[p]
  \center
  \includegraphics[width=0.75\textwidth]{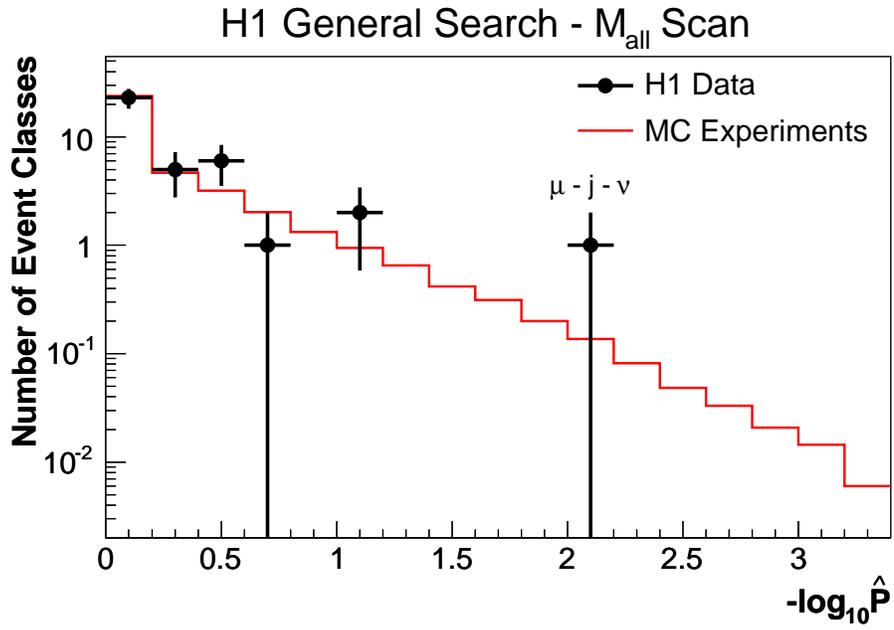}\\[0.5cm]
  \includegraphics[width=0.75\textwidth]{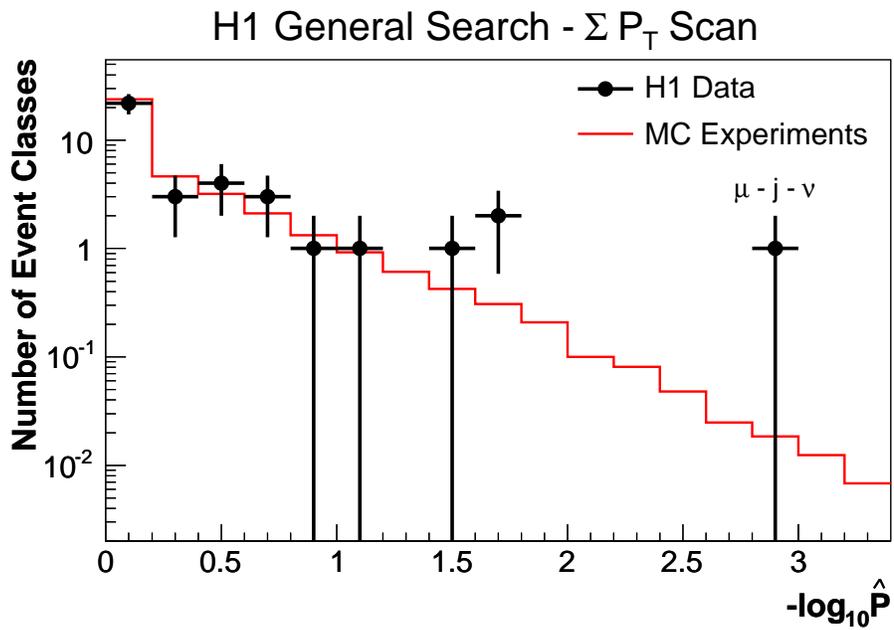}  
  \caption{The $-\log_{10}{\hat{P}}$ values for the data event classes and the 
    expected distribution from MC experiments, as derived by investigating
    the $\Mall$ distributions (top) and $\sum P_T$ distributions
    (bottom) with the search algorithm. 
    } 
  \label{fig:scan}
\end{figure}

The $\hat{P}$ values observed in the data in all event classes are compared in figure~\ref{fig:scan} with
the distribution of $\hat{P}^{SM}$ obtained from the large set of MC experiments, normalised to one experiment. 
The comparison is presented for the scans of the 
$M_{all}$ and $\sum P_T$ distributions.
Most  $\hat{P}$ values lie above $0.01$, corresponding 
to event classes
where no significant discrepancy between the data and the SM expectation 
is observed. 
The global probabilities to find at least
one class with a $\hat{P}$ value smaller than
the observation in the \mujnp~channel
are $3$\% and $28$\% for the $\sum P_T$ and $M_{all}$ distributions,
respectively (see appendix for details).

To test the dependence of the analysis on the {\it a priori} defined 
$P_T$ cuts, the 
whole analysis is repeated with two other object $P_T$ cuts. 
The $P_T$ cut was raised to 
$40$~GeV for all objects and lowered to $15$~GeV. 
In the latter case it
was still required that at least one object has a $P_T$ larger than
$20$~GeV in order to maintain a high trigger efficiency.
The analysis was also repeated 
separately on the $e^+p$ and $e^-p$ data samples. 
In these four test scenarios a similar overall agreement with the SM 
is observed. The \mujnp~event class remains the one with the smallest 
$\hat{P}$ value in the scenario with a lowered $P_T$ cut in the $e^+p$
data sample and no new discrepancy is observed. 
When raising the $P_T$ cut to $40$ GeV, it is mainly the two particle event
classes containing jets that are still populated
and the largest deviation is observed in the \ee~class with $\hat{P}$~=~$0.01$.

\section{Conclusions}
The data collected with the H1 experiment during the years $1994$--$2000$ (HERA I) 
have been investigated in a search 
for deviations from the SM prediction at high transverse 
momentum. 
For the first time all event topologies involving isolated electrons, photons, 
muons, neutrinos and jets are investigated in a single analysis.
A good agreement between the data and the SM expectation is found 
in most event classes. A better knowledge of rare processes 
may be required to search for deviations from the SM in final 
states with four jets at the highest 
invariant mass or transverse momentum.
 The distributions in the
invariant mass and scalar sum of transverse momenta 
of the particles in each event class have been 
systematically searched for 
deviations using a statistical algorithm. 
The most significant deviation is found in the \mujnp~event class, a topology 
where deviations have also been previously reported. 
About $3\%$ ($28\%$) of hypothetical Monte Carlo experiments 
would produce a deviation in at least one event class which is
more significant than that observed 
in the corresponding sum of transverse momenta (invariant mass) 
distribution of the topology with a jet, a muon and a neutrino.

\section*{Acknowledgements}
We are grateful to the HERA machine group whose outstanding efforts have made this 
experiment possible. We thank the engineers and technicians for their work in 
constructing and now maintaining the H1 detector, our funding agencies for 
financial support, the DESY technical staff for continual assistance
and the DESY directorate for support and for the hospitality which they extend to 
the non DESY members of the collaboration. We wish to thank B.~Knuteson for useful
discussions.

\begin{appendix}
\section*{Appendix}
Signals for new physics may appear either as a single deviation or 
a small set of deviations. The following outlines how 
a significant deviation might be defined and presents 
tests of the sensitivity
of this analysis to specific signals for new physics.

The probability $P_X^n$ to observe in the data a $- \log_{10}{\hat{P}}$ 
greater than $X$ in at least $n$ event classes 
is given by the fraction of MC experiments having at least $n$ event classes 
with a $- \log_{10}{\hat{P}^{SM}} > X$.
The $P_X^n$ values obtained for this analysis are presented in 
table~\ref{tab:like}. Up to $3$ event classes are considered. 
Since very similar $P_X^n$ values are found for the $M_{all}$ and 
$\sum P_T$ distributions, averaged values are presented.
For example, a $P_X^n$ value 
smaller than $0.0005$, which might be considered to
represent a significant deviation, could be obtained from
one event class with a 
 $- \log_{10}{\hat{P}} $~$>$~$5$, two event classes with 
a $- \log_{10}{\hat{P}} $~$>$~$3.5$ or three event classes with a
$- \log_{10}{\hat{P}} $~$>$~$3$.
It was found that one of these cases occurs 
either in $M_{all}$ or $\sum P_T$ in
around $0.1\%$ of all MC experiments.

A set of pseudo data samples has been produced to test the sensitivity of the 
analysis procedure to some dedicated signals for new physics. 
The prediction of a specific model for new physics is added to the SM 
prediction and 
this new total prediction is used to generate pseudo data samples.
Again a Monte Carlo technique is used to vary the distribution of signal events and generate many MC experiments.
The complete algorithm is run on those MC experiments 
and the mean value of \mbox{$-\log_{10}{\hat{P}}$} in all of them
is derived as a measure of sensitivity of this analysis. 

\begin{figure}[p]
  \center
  \includegraphics[width=0.5\textheight]{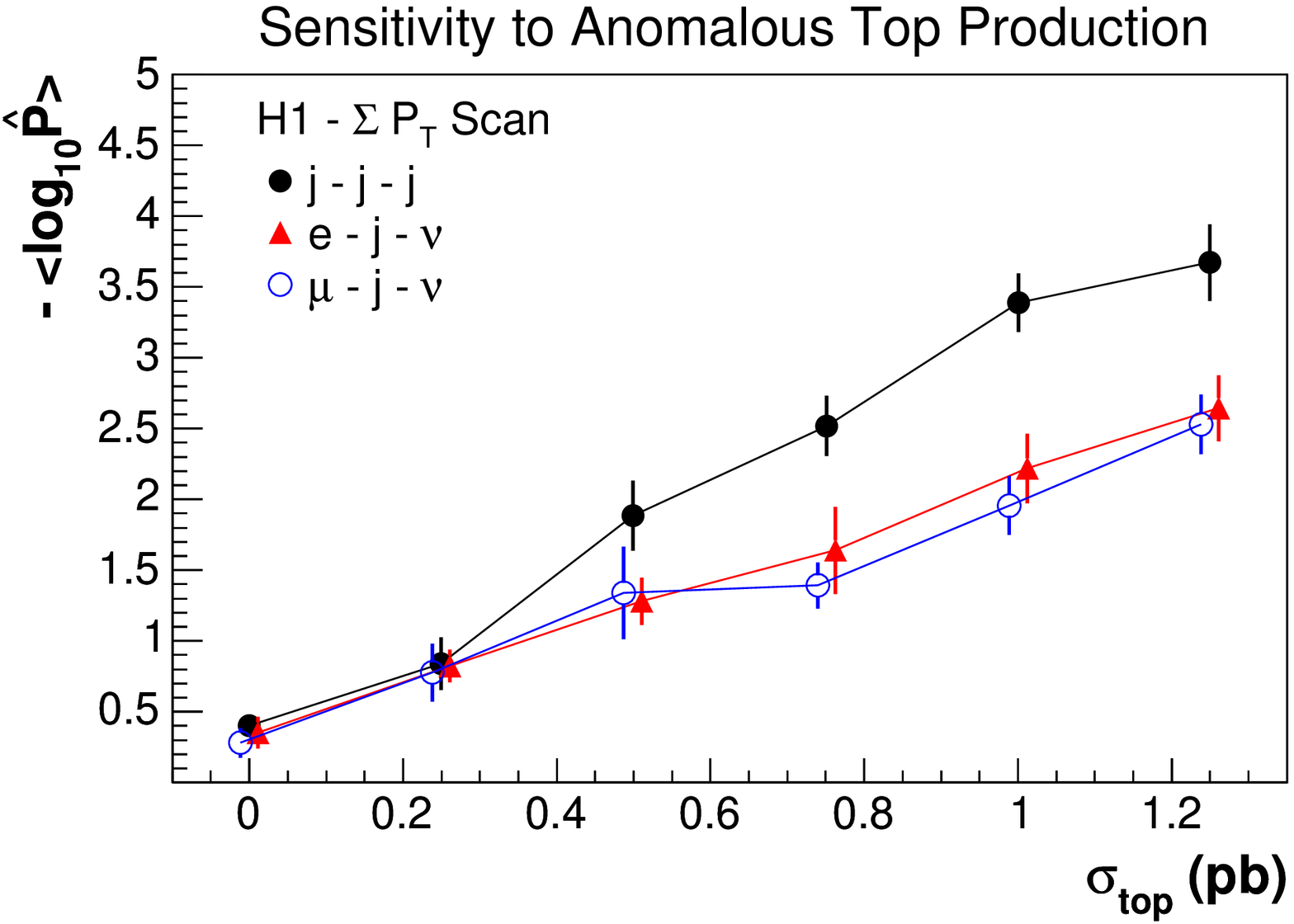}
  \includegraphics[width=0.5\textheight]{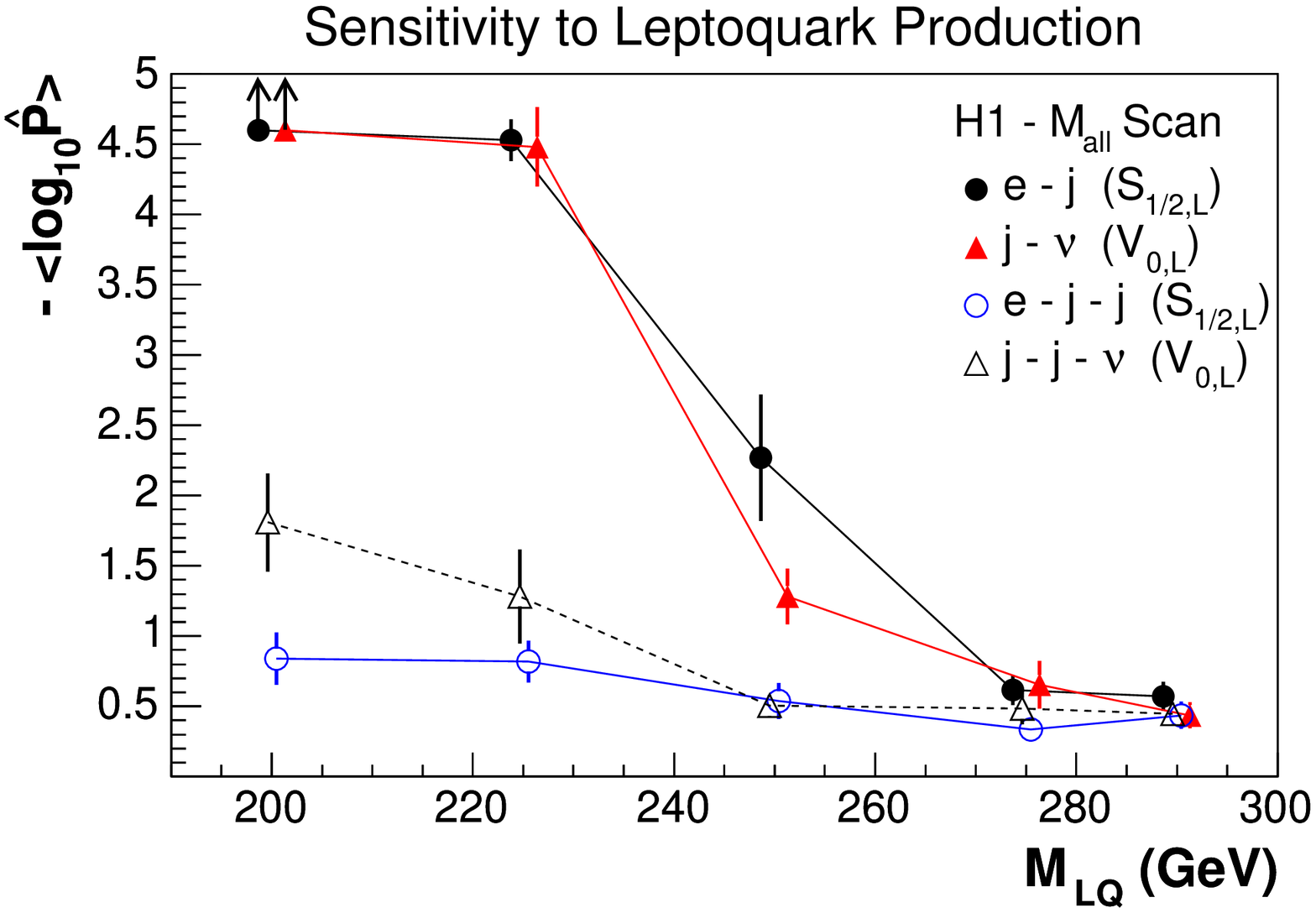}
  \caption{The mean value of $-\log_{10}{\hat{P}}$ as derived from MC experiments
    which include a top signal with a cross section $\sigma_{top}$ (top) and 
    a LQ signal with a mass $M_{LQ}$ and a  $\lambda$ coupling equal to $0.05$ (bottom), using the distributions of $\sum P_T$
    and $M_{all}$, respectively. The small arrows on the bottom figure indicate that these values should be treated as lower limits. }
  \label{fig:alpha}
\end{figure}

The exotic production of top quarks via a flavour-changing 
neutral current is first investigated. 
The decay $t \rightarrow b W$ with subsequent leptonic and
hadronic $W$ decays has been considered.
The \mbox{$\langle -\log_{10}{\hat{P}} \rangle$} values obtained are displayed in 
figure~\ref{fig:alpha} (top) as a function of the cross section
for producing a top when the proton beam energy is 920 GeV.
Whereas \mbox{$\langle -\log_{10}{\hat{P}} \rangle$} is around $0.43$ if no signal is present, 
the value increases if a top is produced. 
In the \jjj~event class a \mbox{$\langle -\log_{10}{\hat{P}} \rangle$} of $2$ is obtained for a cross-section $\sigma_{top}$ of $\sim$~$0.5$~pb. 
This value can be compared with the $95$\% confidence level exclusion limit 
on the top production cross section at $\sigma_{top} < 0.48$~pb 
already derived 
by the H1 experiment using the hadronic top decay channel only~\cite{Aktas:2003yd}. A deviation with three event classes with a
$\langle - \log_{10}{\hat{P}} \rangle$~$>$~$3$ would be 
found in this example for 
$\sigma_{top} \approx 1.5$~pb.
 
The second test concerns the production of leptoquarks (LQs)~\cite{Adloff:1999tp}. 
$S_{1/2,L}$ and $V_{0,L}$ type leptoquarks have been considered, which would 
mainly manifest themselves 
in the \ej~and \jnp~channels. A $\lambda$ coupling of $0.05$ has been 
assumed and the sensitivity 
of the present analysis was tested for various LQ masses. The 
$\langle - \log_{10}{\hat{P}} \rangle$ values obtained 
from searches in the $M_{all}$ distributions are summarised in 
figure~\ref{fig:alpha} (bottom), 
for both the $S_{1/2,L}$ and $V_{0,L}$ LQ appearing 
in the \ej~and~\ejj~as well as the~\jnp~and~\jjnp~channels,
respectively. This analysis is sensitive to both types of 
leptoquarks up to masses of $240-250$~GeV.
These values can be compared with $95$\% confidence level limits of
 $265$~GeV for $S_{1/2,L}$ LQs 
and $240$~GeV for $V_{0,L}$ LQs, determined by dedicated analyses~\cite{Chekanov:2003af}. As for the case of single top production, 
the general search is thus found to have 
a sensitivity to leptoquark production which is
comparable with that of dedicated searches.

\end{appendix}

\clearpage

\begin{table}[b]
  \begin{center}
    \begin{tabular}{|l|c|c|c|}
      \hline
      Object & Energy Scale & $\theta$ &  Identification \\
      &          &    (mrad)    &     efficiency \\ 
      \hline                                        
      Jet &  2\%             &  5--10   &  -- \\ 
      Electron & 0.7--3\%    &  1--3    &  2--7\%   \\    
      Photon & 0.7--3\%      &  1--3    &  2--7\% \\ 
      Muon &  5\%            &  3       &  5\% \\ 
      \hline
    \end{tabular}
    \caption{Systematic uncertainties attributed to the measurement of 
energies and polar angles and to the identification efficiencies of particles.}
    \label{tab:objectunc}
  \end{center}  
\end{table}

\begin{table}[b]
  \begin{center}
    \begin{tabular}{|l|c|}
      \hline
      Process &Uncertainty\\
      \hline
      $ep\rightarrow jj X$ and $ep \rightarrow j\gamma X$ &15\% \\
      $ep\rightarrow j\nu X$ and $ep \rightarrow jeX$  & 10\% \\
      $ep \rightarrow jj\nu X$ and $ep \rightarrow jjeX$  & 15\% \\
      $ep \rightarrow \mu\mu X$ and $ep \rightarrow ee X$ & 3\%\\
      $ep \rightarrow W X$ and $ep \rightarrow WjX$ & 15\%\\
      $ep \rightarrow e\gamma X$ and $ep \rightarrow e\gamma j$& 10\% \\
      $ep \rightarrow e\gamma p$ & 5\% \\
      \hline
    \end{tabular}
    \caption{Theoretical uncertainties attributed to the simulation of different SM processes.}
    \label{tab:modelunc}
  \end{center}
\end{table}

\begin{table}[p]
  \scriptsize{
    \begin{center} 
      \begin{tabular}{|c||c|r|rcr|c||c|r|rcr|c|} 
        \hline
        & \multicolumn{6}{c||}{$M_{all}$} & \multicolumn{6}{c|}{$\sum P_T$} \\
        \hline
        \hline
        event class & $\hat{P}$ & $N_{obs}$ & 
        $N_{SM}$ & $\pm$ & $\delta 
        N_{SM}$ & $p$ 
        & $\hat{P}$ & $N_{obs}$ & 
        $N_{SM}$ & $\pm$ & $\delta 
        N_{SM}$ & $p$\\
        \hline
        \jj &  0.38    &   1  &  0.035  & $\pm$ &  0.017  &  0.036  &      0.12     &   1  &  0.013  & $\pm$ &  0.006  &  0.013\\
        \ej &  0.94    & 111  &  139  & $\pm$ &  21  &  0.12 &    0.021     &  12  &  31.2  & $\pm$ &  5.1  &  0.0028\\
        \muj &  0.67    &   3  &  1.07  & $\pm$ &  0.25  &  0.098 &       0.29     &   3  &  0.70  & $\pm$ &  0.23  &  0.040\\
        \jnp &  0.34    &  83  &  116  & $\pm$ &  14  &  0.028 &    0.22     &  20  &  36.7  & $\pm$ &  6.2  &  0.023\\
        \enp &  0.94    &   5  &  10.6  & $\pm$ &  4.4  &  0.17 &      0.77     &   0  &  2.1  & $\pm$ &  0.8  &  0.17\\
        \ee &  0.32    &   3  &  0.56  & $\pm$ &  0.17  &  0.023 &       0.019     &   3  &  0.18  & $\pm$ &  0.08  &  0.0013\\
        \emu &  0.21    &   4  &  0.93  & $\pm$ &  0.12  &  0.016 &       0.56     &   0  &  2.6  & $\pm$ &  0.5  &  0.080\\
        \mumu &  0.069    &   2  &  0.14  & $\pm$ &  0.04  &  0.010 &       0.036     &   2  &  0.11  & $\pm$ &  0.03  &  0.0060\\
        \jpho &  0.52    &   3  &  10.8  & $\pm$ &  3.7  &  0.052 &      0.77     &   0  &  2.5  & $\pm$ &  1.0  &  0.13\\
        \epho &  0.38    &   9  &  19.2  & $\pm$ &  2.0  &  0.014 &      0.64     &   8  &  15.7  & $\pm$ &  1.9  &  0.040\\
        \phopho &  0.47    &   1  &  0.16  & $\pm$ &  0.09  &  0.15 &       0.31     &   1  &  0.11  & $\pm$ &  0.09  &  0.12\\
        \jjj &  0.41   &  12  &  5.9  & $\pm$ &  2.0  &  0.050 &       0.58    &  14  &  7.8  & $\pm$ &  2.5  &  0.077\\
        \ejj &  0.69   &  39  &  59.6  & $\pm$ &  10.7  &  0.058 &     0.085    &   9  &  23.9  & $\pm$ &  4.4  &  0.0072\\
        \jjnp &  0.62   &   5  &  1.79  & $\pm$ &  0.41  &  0.043 &       0.51    &   5  &  1.74  & $\pm$ &  0.45  &  0.040\\
        \ejnp &  0.090   &   2  &  0.19  & $\pm$ &  0.05  &  0.016 &       0.16    &   2  &  0.28  & $\pm$ &  0.06  &  0.034\\
        \mujnp &  $9.7 \cdot 10^{-3}$  &   3  &  0.19  & $\pm$ &  0.05  &  0.0011 &       $1.0 \cdot 10^{-3}$    &   3  &  0.068  & $\pm$ &  0.029  &  $7.5 \cdot 10^{-5}$\\
        \jjpho &  0.27   &   1  &  0.074  & $\pm$ &  0.048  &  0.076 &       0.36    &   1  &  0.15  & $\pm$ &  0.10  &  0.15\\
        \ejpho &  0.47   &   1  &  5.7  & $\pm$ &  1.6  &  0.050 &       0.39    &   1  &  5.6  & $\pm$ &  1.4  &  0.045\\
        \ejjj &  0.98  &   0  &  1.6  & $\pm$ &  0.5  &  0.23 &       0.87   &   1  &  0.18  & $\pm$ &  0.06  &  0.17 \\
        \jjjnp &  0.33  &   1  &  0.084  & $\pm$ &  0.045  &  0.083 &       0.20   &   2  &  0.31  & $\pm$ &  0.14  &  0.044 \\
        \hline
      \end{tabular} 
    \end{center} 
  }
  \caption{The $\hat{P}$ values, the number of data events $N_{obs}$ and the 
    SM expectation $N_{SM}$ for the region  
    derived by the search algorithm using the $M_{all}$ and $\sum P_T$ distributions 
    for event classes containing at least one event and taken into account in the statistical procedure. The $p$ value in the selected region is also presented. 
}
  \label{tab:phatmass}
\end{table}


\begin{table}[b]
  \begin{center}
    \begin{tabular}{|l|c|c|c|c|c|c|c|c|c|}
      \hline
      $n$ & $P_1$ & $P_{1.5}$ & $P_2$ & $P_{2.5}$ & $P_3$ & $P_{3.5}$ & $P_4$ & $P_{4.5}$ & $P_5$ \\
      \hline                                        
      1 & 95\% & 65\%  & 28\%  & 9\%    & 3\%       & 0.9\%       & 0.2\%     & 0.1\%    & $<$0.05\% \\ 
      2 & 79\% & 28\%  & 4\%   & 0.6\%  & 0.1\%     & $<$0.05\% & --- & --- & --- \\    
      3 & 53\% & 8\%   & 0.4\% & 0.05\% & $<$0.05\% & ---       &  ---      & ---       & --- \\ 
      \hline
    \end{tabular}
    \caption{The probability $P_X^n$ to find at least $n$ event classes with
      a \mbox{$- \log{\hat{P}} $} value greater than $X$. 
      The values are applicable to both the $\Mall$ and $\sum P_T$ analyses.
    }
    \label{tab:like}
  \end{center}  
\end{table}

\clearpage

\clearpage

\clearpage

\clearpage

\end{document}

%% file: h1auts.tex

A.~Aktas$^{10}$,               
V.~Andreev$^{26}$,             
T.~Anthonis$^{4}$,             
A.~Asmone$^{33}$,              
A.~Babaev$^{25}$,              
S.~Backovic$^{37}$,            
J.~B\"ahr$^{37}$,              
P.~Baranov$^{26}$,             
E.~Barrelet$^{30}$,            
W.~Bartel$^{10}$
% ...

\bigskip{\it
% ...
$ ^{4}$ Inter-University Institute for High Energies ULB-VUB, Brussels;
          Universiteit Antwerpen, Antwerpen; Belgium$^{ c}$ \\
% ...
$ ^{10}$ DESY, Hamburg, Germany \\
% ...
$ ^{25}$ Institute for Theoretical and Experimental Physics,
          Moscow, Russia$^{ l}$ \\
 $ ^{26}$ Lebedev Physical Institute, Moscow, Russia$^{ e}$ \\
% ...
$ ^{30}$ LPNHE, Universit\'{e}s Paris VI and VII, IN2P3-CNRS,
          Paris, France \\
% ...
$ ^{33}$ Dipartimento di Fisica Universit\`a di Roma Tre
          and INFN Roma~3, Roma, Italy \\
% ...
$ ^{37}$ DESY, Zeuthen, Germany \\
% ...
$ ^c$ Supported by FNRS-FWO-Vlaanderen, IISN-IIKW and IWT
      and  by Interuniversity Attraction Poles Programme,
      Belgian Science Policy \\
% ...
$ ^e$ Supported by the Deutsche Forschungsgemeinschaft \\
% ...
$ ^l$ Partially Supported by Russian Foundation
      for Basic Research, grant    no. 00-15-96584 \\
}

%% file: submit.bbl
\begin{thebibliography}{99}
\bibitem{Kuze:2002vb}
M.~Kuze and Y.~Sirois,
Prog.\ Part.\ Nucl.\ Phys.\  {\bf 50} (2003) 1
[hep-ex/0211048].

\bibitem{Abbott:2001ke}
B.~Abbott {\it et al.}  [D0 Collaboration],
Phys.\ Rev.\ D {\bf 62} (2000) 092004
[hep-ex/0006011].  \\
B.~Abbott {\it et al.}  [D0 Collaboration],
Phys.\ Rev.\ D {\bf 64} (2001) 012004
[hep-ex/0011067].

\bibitem{Brun:1987ma}
R.~Brun, F.~Bruyant, M.~Maire, A.~C.~McPherson and P.~Zanarini,
CERN-DD/EE/84-1.

\bibitem{Sjostrand:2000wi}
T.~Sj\"{o}strand  {\it et al.},
Comput.\ Phys.\ Commun.\  {\bf 135} (2001) 238
[hep-ph/0010017].\\
(The PARP(67) parameter was set to 4 instead of its default value of 1).

\bibitem{Jung:1993gf}
H.~Jung,
Comput.\ Phys.\ Commun.\  {\bf 86} (1995) 147.

\bibitem{Kwiatkowski:1990es}
A.~Kwiatkowski, H.~Spiesberger and H.~J.~M\"{o}hring,
Comput.\ Phys.\ Commun.\  {\bf 69} (1992) 155.

\bibitem{Schuler:yg}
G.~A.~Schuler and H.~Spiesberger,
``Django: The Interface for The Event Generators Heracles and Lepto''.

\bibitem{Lonnblad:1992tz}
L.~L\"{o}nnblad,
Comput.\ Phys.\ Commun.\  {\bf 71} (1992) 15.

\bibitem{Berger:kp}
C.~Berger and P.~Kandel,
Prepared for Workshop on Monte Carlo Generators for HERA Physics Hamburg, Germany, 27-30 Apr 1998.

\bibitem{Abe:2000cv}
T.~Abe,
Comput.\ Phys.\ Commun.\  {\bf 136} (2001) 126
[hep-ph/0012029].

\bibitem{Mucke:1999yb}
A.~Mucke  {\it et al.},
Comput.\ Phys.\ Commun.\  {\bf 124} (2000) 290
[astro-ph/9903478].

\bibitem{Baur:1991pp}
U.~Baur, J.~A.~Vermaseren and D.~Zeppenfeld,
Nucl.\ Phys.\ B {\bf 375} (1992) 3.

\bibitem{Diener:2002if}
K.~P.~Diener, C.~Schwanenberger and M.~Spira,
Eur.\ Phys.\ J.\ C {\bf 25} (2002) 405
[hep-ph/0203269].

\bibitem{Diener:2003df}
K.~P.~Diener, C.~Schwanenberger and M.~Spira,
[hep-ex/0302040].

\bibitem{Adloff:2002au}
C.~Adloff {\it et al.}  [H1 Collaboration],
Eur.\ Phys.\ J.\ C {\bf 25} (2002) 13
[hep-ex/0201006].

 
\bibitem{Abt:1996xv}
I.~Abt {\it et al.}  [H1 Collaboration],
Nucl.\ Instrum.\ Meth.\ A {\bf 386} (1997) 348;\\
I.~Abt {\it et al.}  [H1 Collaboration],
Nucl.\ Instrum.\ Meth.\ A {\bf 386} (1997) 310.

\bibitem{Andrieu:1993kh}
B.~Andrieu {\it et al.}  [H1 Calorimeter Group Collaboration],
Nucl.\ Instrum.\ Meth.\ A {\bf 336} (1993) 460.
 
\bibitem{Appuhn:1996na}
R.~D.~Appuhn {\it et al.}  [H1 SPACAL Group Collaboration],
Nucl.\ Instrum.\ Meth.\ A {\bf 386} (1997) 397.

\bibitem{Adloff:2003uh}
C.~Adloff {\it et al.}  [H1 Collaboration],
Eur.\ Phys.\ J.\ C {\bf 30}, 1 (2003)
[hep-ex/0304003].


\bibitem{matti}
M.~Peez, ``Recherche de d\'eviations au Model Standard dans les processus de grande \'energie 
transverse sur le collisionneur \'electron - proton HERA'', Ph.D. thesis, Universit\'e de Lyon (2003), DESY-THESIS-2003-023.
(available at http://www-h1.desy.de/psfiles/theses/).

\bibitem{Aktas:2003sz}
  A.~Aktas {\it et al.}  [H1 Collaboration],
Phys.\ Lett.\ B {\bf 583} (2004) 28
[hep-ex/0311015].


\bibitem{Zhang:2000hb}
Z.~Zhang,  Habilitation Thesis, LAL preprint, LAL 00-57 (2000)
[hep-ph/0012249].



\bibitem{Andreev:2003pm}
V.~Andreev {\it et al.}  [H1 Collaboration],
Phys.\ Lett.\ B {\bf 561} (2003) 241
[hep-ex/0301030].

\bibitem{Ellis:1993tq}
S.~D.~Ellis and D.~E.~Soper,
Phys.\ Rev.\ D {\bf 48} (1993) 3160
[hep-ph/9305266].
 
\bibitem{Catani:1993hr}
S.~Catani, Y.~L.~Dokshitzer, M.~H.~Seymour and B.~R.~Webber,
Nucl.\ Phys.\ B {\bf 406} (1993) 187.



\bibitem{Giele:1997hd}
W.~T.~Giele, E.~W.~N.~Glover and D.~A.~Kosower,
Phys.\ Rev.\ D {\bf 57} (1998) 1878
[hep-ph/9706210].


\bibitem{frising}
G. Frising, ``Rare Phenomena and $W$ production in Electron-Proton
Scattering at HERA'', Ph.D. thesis, RWTH Aachen (2003)
(available at http://www-h1.desy.de/psfiles/theses/).

\bibitem{martin}
M. Wessels, ``General Search for New Phenomena in ep Scattering at HERA'', Ph.D. thesis, RWTH Aachen (2004)
(available at http://www-h1.desy.de/psfiles/theses/).




\bibitem{Aktas:2003jg}
A.~Aktas {\it et al.}  [H1 Collaboration],
Eur.\ Phys.\ J.\ C {\bf 31} (2003) 17
[hep-ex/0307015].

\bibitem{Aktas:2003yd}
A.~Aktas {\it et al.}  [H1 Collaboration],
Eur.\ Phys.\ J.\ C {\bf 33}, 9 (2004)
[hep-ex/0310032].




\bibitem{Adloff:1999tp}
C.~Adloff {\it et al.}  [H1 Collaboration],
Eur.\ Phys.\ J.\ C {\bf 11} (1999) 447
[Erratum-ibid.\ C {\bf 14} (2000) 553]
[hep-ex/9907002].


%


\bibitem{Chekanov:2003af}
S.~Chekanov {\it et al.}  [ZEUS Collaboration],
Phys.\ Rev.\ D {\bf 68} (2003) 052004
[hep-ex/0304008].




\end{thebibliography}
